\documentstyle[12pt]{article}
\def\be{\begin{equation}}
\def\ee{\end{equation}}
\def\ba{\begin{eqnarray}}
\def\ea{\end{eqnarray}}

\def\fun#1#2{\lower3.6pt\vbox{\baselineskip0pt\lineskip.9pt

\ialign{$\mathsurround=0pt#1\hfill##\hfil$\crcr#2\crcr\sim\crcr}}}
\begin{document}

\begin{titlepage}
\null\vspace{-62pt}
\begin{flushright} UMD-PP-95-143\\
 hep-ph/9506471 \\
    June 1995
\end{flushright}
\vspace{0.2in}

\centerline{{\large \bf  Spontaneous CP violation in supersymmetric}}

\centerline{{\large \bf  models with four Higgs doublets} }

\vspace{0.5in}
\centerline{
Manuel Masip$^{(a,b)}$ \ \ and \ \ Andrija Ra\v{s}in$^{(a)}$ }
\vspace{0.2in}
\centerline{\it $^{(a)}$Department of Physics}
\centerline{\it University of Maryland}
\centerline{\it College Park, MD 20742, U.S.A.}
\vspace{.2in}
\centerline{\it $^{(b)}$Departamento de F\'\i sica
Te\'{o}rica y del Cosmos}
\centerline{\it Universidad de Granada}
\centerline{\it 18071 Granada, Spain}
\vspace{.7in}
\baselineskip=22pt

\centerline{\bf Abstract}
\begin{quotation}

We consider supersymmetric extensions of the standard model
with two pairs of Higgs doublets. We study the possibility
that CP violation is generated spontaneously in the scalar
sector via vacuum expectation values (VEVs) of the Higgs
fields. Using a simple geometrical interpretation of the
minimum conditions we prove that the minimum of the tree-level scalar
potential for these models is allways real. We show that
complex VEVs can appear once radiative corrections and/or
explicit {\it soft} CP violating terms are added to the
effective potential.

\vspace{0.4in}

\end{quotation}
\end{titlepage}

\baselineskip=24pt

\section{Introduction}

In order to introduce CP violation into gauge models
one can consider two different approaches \cite{moha92}.
CP violating phases could
be intrinsic to the parameters of the original Lagrangian or,
alternatively,
they could be spontaneous in the sense that all parameters of
the theory are real, but the vacuum expectation values (VEVs)
of the scalar Higgs fields are complex.
Experimental evidence shows that CP violation is {\it small}
but nonzero (it has only been measured in kaon physics). In
general, this
fact provides a {\it naturalness} criterium that can be
used to decide which approach to generate CP violation
seems favoured for a definite
model. The first approach will be more natural ({\it i.e.},
require less fine tuned idependent parameters) in models
where after phase redefinitions of the fields one is
left with few independent phases, whereas spontaneous CP
violation (SCPV) seems preferred for models with a large
number of parameters and a rich enough scalar sector.

In the standard model, for example, assuming the
parameters in the Lagrangian complex, all phases can be
absorbed by field redefinitions except for two: the QCD
phase $\theta$ and the phase $\phi$ in the
Cabibbo-Kobayashi-Maskawa (CKM) matrix. The most popular
mechanism proposed  to solve the strong CP problem
(that we shall not treat here) requires the introduction
of a global symmetry \cite{pecc77}, whereas the CKM phase appears allways
multiplied by the small mixing $V_{13}$ and do not require
a fine tuned value to explain the kaon system. In this
case SCPV, possible only by extending the Higgs
and/or fermion
sectors, is less economical.
In contrast, in supersymmetric (SUSY) extensions of the standard
model the origin of CP violation is more involving,
due essentially to the large number of soft SUSY breaking
parameters in the Lagrangian.
For arbitrary complex parameters there are several
new sources of CP violation, and the prediction of a neutron electric
dipole moment within the experimental limits requires cancellations
of two or three orders of magnitude. In this framework, the
idea of SCPV seems appealing. Unfortunately,
it is known that in the minimal SUSY extension of the standard
model (MSSM) there is no spontaneous generation of
CP violation\footnote{In fact, SCPV in MSSM is in
principle possible once the
radiative corrections are included \cite{maek92}. However, this
model contains a too light boson, and is thus experimentally
excluded \cite{poma92}.}.

The first SUSY extension where SCPV has been found is in models
with gauge
singlets \cite{roma86,poma93}, although in the simplest case
the complex VEVs appear only for one-loop effective
potentials with very strong top corrections.
The addition of singlets does not spoil the
gauge unification of the MSSM, relaxes bounds on the
mass of the lightest neutral scalar, and offers a possible explanation
to the $\mu$ problem (the Higgs mass term in the superpotential).
Moreover, as shown by Pomarol \cite{poma93}, the phases generated
spontaneously in these models could be enough to explain
CP violation in kaon physics ({\it i.e.}, no complex Yukawas
would be necessary).
The singlet models, however, seem to imply either the presence of
light Higgs scalars and heavy squarks or small complex phases,
which in turn invoke some degree of fine tuning.

In this article we study the possibility of SCPV in
SUSY models with two pairs of Higgs doublets. These models are
an obvious generalization of the minimal case (they do not introduce
new species, just {\it double} the Higgs sector). They occur
naturally in left-right symmetric scenarios, where at least two
bidoublets are required in order to get realistic fermion masses
and mixings. It was also shown \cite{nels93} that (unlike
the minimal or singlet SUSY models) four Higgs doublet
models can have a large $\tan\beta$ without fine
tuning or too light charginos.
Like in the singlet case, the
lightest scalar in the four Higgs doublet models
has {\it not} necessarily a tree-level mass smaller
than the $Z$ mass. The fact that four doublet models require an
intermediate scale to be consistent with gauge unification
could also be an advantage \cite{brah95}, since it might
be more in line with recent data on $\alpha_s(M_Z)$
than minimal unification scenarios.

Since more than one Higgs doublet couples to quarks of a given charge,
these models have in principle too large
flavor changing neutral currents (FCNC)
and  require a mechanism  to suppress some
Yukawa couplings
(usually, an approximate flavor symmetry \cite{anta92}).
If the couplings are complex with phases of order one
these models also tend to have too large
CP violation in
the kaon system and too large neutron electric dipole moment
\cite{hall93}, requiring an additional mechanism to suppress
phases or couplings.
These questions, briefly addressed in this paper, will be
studied in detail elsewhere in the framework
of models with approximate global
symmetries \cite{rasi95,masi95}.

\section{The Higgs sector and the minimum conditions}

We start defining the model and establishing the conditions
for the minimum of the tree-level potential.
For previous work on
SUSY models with four Higgs doublets see
for example \cite{habe90}.

The Higgs sector of the model contains two pairs of $SU(2)$ doublet
superfields, $(H_1,H_3)$ and $(H_2,H_4)$, with hypercharges $-1$ and
$+1$, respectively. We denote these doublets by
\be
H_{1(3)} =
\left(
\begin{array}{c}
\phi^0_{1(3)}  \\
\phi^-_{1(3)}
\end{array}
\right), \;\;
H_{2(4)} =
\left(
\begin{array}{c}
\phi^+_{2(4)}  \\
\phi^0_{2(4)}
\end{array}
\right).
\ee
The most general superpotential with four higgs doublets
is then given by
\ba
W & = & Q ( {\bf h}_1 H_1 + {\bf h}_3 H_3) D^c +
Q ( {\bf h}_2 H_2 + {\bf h}_4 H_4) U^c +
        L({\bf h}_1^e H_1 + {\bf h}_3^e H_3) E^c \nonumber\\
& + & \mu_{12} H_1 H_2 + \mu_{32} H_3 H_2
+ \mu_{14} H_1 H_4 + \mu_{34} H_3 H_4 ,
\label{eq:superpot}
\ea
where $Q$ stand for quark doublets, $D^c$ for down quark singlets,
$U^c$ for up quark singlets, $L$ for lepton doublets, $E^c$ for
charged lepton singlets, and  ${\bf h}_i$ are the Yukawa matrices
(family indices are omitted).

Including soft SUSY breaking terms,
the most general tree level scalar potential involving only Higgs fields
is given by
\ba
V & = & m_1^2 H_1^\dagger H_1+ m_2^2 H_2^\dagger H_2
 + m_3^2 H_3^\dagger H_3
+ m_4^2 H_4^\dagger H_4 + \nonumber\\
 & + & (m^2_{12} H_1 H_2 + h.c.) + (m^2_{32} H_3 H_2 + h.c.) + \nonumber\\
 & + & (m^2_{14} H_1 H_4 + h.c.) + (m^2_{34} H_3 H_4 + h.c.) + \nonumber\\
 & + & (m^2_{13} H_1^\dagger H_3 + h.c.)
+ (m^2_{24} H_2^\dagger H_4 + h.c.) + V^{4HD}_D ,
\label{eq:pot}
\ea
where $V^{4HD}_D$ is the D-term part of the potential. For the
neutral components of the doublets one has
\ba
V^{4HD}_D & = & {1 \over 8} (g^2+g'^2) [
\phi^{0\;\dagger}_1 \phi^{0}_1 +
\phi^{0\;\dagger}_3 \phi^{0}_3 -
\phi^{0\;\dagger}_2 \phi^{0}_2 -
\phi^{0\;\dagger}_4 \phi^{0}_4 ]^2 .
\label{eq:dpot}
\ea

We will assume that the theory is CP invariant, {\it i.e.},
all the couplings and mass parameters are real.
We do not assume any higher energy scales
or accidental cancellations, but we suppose
that there is a part of the parameter space
giving minima of the potential
which do not break the electric charge.

After spontaneous symmetry
breaking, the Higgs fields will acquire VEVs
which are possibly complex (from now on we drop the 0 superscript
to specify neutral fields):
\be
<\phi_1> = {1 \over \sqrt{2} } v_1  \; ; \;\;
<\phi_3> = {1 \over \sqrt{2} } v_3 e^{i\delta_3} \, ,
\ee
and
\be
<\phi_2> = {1 \over \sqrt{2} } v_2 e^{i\delta_2}  \; ; \;\;
<\phi_4> = {1 \over \sqrt{2} } v_4 e^{i\delta_4} \, ,
\ee
where we have used a global hypercharge transformation to
rotate away the phase of $<\phi_1>$.
The vacuum expectation value of the scalar potential is then
\ba
< V > & = &{1 \over 2} m_1^2 v_1^2 +
{1 \over 2} m_2^2 v_2^2 +
{1 \over 2} m_3^2 v_3^2 +
{1 \over 2} m_4^2 v_4^2 +
m_{12}^2 v_1 v_2 \cos \delta_2 + m_{13}^2 v_1 v_3 \cos \delta_3  +
 \nonumber \\
& + &  m_{14}^2 v_1 v_4 \cos \delta_4
+ m_{32}^2 v_3 v_2 \cos(\delta_3+\delta_2)
+ m_{34}^2 v_3 v_4 \cos(\delta_3+\delta_4)
+  \nonumber \\
& + & m_{24}^2 v_2 v_4 \cos(\delta_2-\delta_4) +
{1 \over 32} (g^2 + g'^2) [ v_1^2 + v_3^2 - v_2^2 - v_4^2 ]^2 .
\label{eq:vevpot}
\ea
The conditions at the minimum are
\ba
\frac {\partial V} {\partial v_1} & = &
    m_1^2 v_1 + m_{12}^2 v_2 \cos \delta_2
    + m_{13}^2 v_3 \cos \delta_3 + m_{14}^2 v_4 \cos \delta_4
    + v_1 g({\bf v}) = 0 \, , \nonumber\\
\frac {\partial V} {\partial v_2} & = &
    m_2^2 v_2 + m_{12}^2 v_1 \cos \delta_2
    + m_{32}^2 v_3 \cos (\delta_3 + \delta_2)
    + m_{24}^2 v_4 \cos (\delta_2 - \delta_4)
    - v_2 g({\bf v}) = 0 \, , \nonumber\\
\frac {\partial V} {\partial v_3} & = &
    m_3^2 v_3 + m_{32}^2 v_2 \cos (\delta_3 + \delta_2)
    + m_{13}^2 v_1 \cos \delta_3
    + m_{34}^2 v_4 \cos (\delta_3 + \delta_4)
    + v_3 g({\bf v}) = 0 \, , \nonumber\\
\frac {\partial V} {\partial v_4} & = &
    m_4^2 v_4  + m_{24}^2 v_2 \cos (\delta_2 - \delta_4)
    + m_{34}^2 v_3 \cos (\delta_3 + \delta_4)
    + m_{14}^2 v_1 \cos \delta_4
    - v_4 g({\bf v}) = 0
\label{eq:minvevs}
\ea
where $g({\bf v}) =   {1 \over 8} (g^2 + g'^2)
            [ v_1^2 + v_3^2 - v_2^2 - v_4^2 ]$,
and
\ba
\frac {\partial V} {\partial \delta_2} & = &
    m_{12}^2 v_1 v_2 \sin \delta_2
    + m_{32}^2 v_3 v_2 \sin (\delta_3 + \delta_2)
    + m_{24}^2 v_2 v_4 \sin (\delta_2 - \delta_4) = 0 \, , \nonumber\\
\frac {\partial V} {\partial \delta_3} & = &
     m_{32}^2 v_3 v_2 \sin (\delta_3 + \delta_2)
    + m_{13}^2 v_1 v_3 \sin \delta_3
    + m_{34}^2 v_3 v_4 \sin (\delta_3 + \delta_4) = 0 \, , \nonumber\\
\frac {\partial V} {\partial \delta_4} & = &
    - m_{24}^2 v_2 v_4 \sin (\delta_2 - \delta_4)
    + m_{34}^2 v_3 v_4 \sin (\delta_3 + \delta_4)
    + m_{14}^2 v_1 v_4 \sin \delta_4 = 0 .
\label{eq:minphases}
\ea

The seven equations in (\ref{eq:minvevs})
and (\ref{eq:minphases}) contain seven unknows: the four
VEVs $v_i$ ($i=1,...,4$) and the three phases $\delta_i$
($i=2,3,4$), and thus can be in principle solved.
The last three equations have a trivial
solution where all the sines vanish.
However, a necessary condition for the theory to have spontaneous CP
violation is that at least one of
the phases $\delta_i$ is different
from $0$ or $\pi$.
Therefore we are looking for nontrivial solutions to (\ref{eq:minphases}).
The easiest way to solve this problem is to
realize that
there is a geometrical object that
satisfies these equations\footnote{
This is similar to most searches for SCPV in models with
two phases. In that case, the nontrivial solution is found
when the two phases can be fit as angles in a triangle with
sides related to the VEVs and mass parameters
in the potential \cite{bran80}.}.
It is defined by
3 triangles, each of which has two of the angles $\delta_i$
as shown in Figure 1.
Appropriately, we call this object ``tri-triangle".
Six of the nine sides of the tri-triangle are independent.
Addition of the sine laws of the three triangles
gives
\ba
         (a - x) \sin \delta_2
    & + &   c    \sin (\delta_3 + \delta_2)
      \, + \,   d  \sin (\delta_2 - \delta_4) = 0 \, , \nonumber\\
            c    \sin (\delta_3 + \delta_2)
    & + &(b - z) \sin \delta_3
      \, +  \,  f    \sin (\delta_3 + \delta_4) = 0 \, , \nonumber\\
      -     d    \sin (\delta_2 - \delta_4)
    & + &   f    \sin (\delta_3 + \delta_4)
      \, + \, (e - y) \sin \delta_4  = 0 \, .
\label{eq:tritphase}
\ea
Comparing (\ref{eq:minphases}) and (\ref{eq:tritphase}) we find
the correspondence between the six independent distances in the tri-triangle
and the six independent quantities $m^2_{ij}v_iv_j$:
\ba
a-x = m_{12}^2 v_1 v_2 &,& c = m_{32}^2 v_2 v_3 \, , \nonumber\\
b-z = m_{13}^2 v_1 v_3 &,& d = m_{24}^2 v_2 v_4 \, , \nonumber\\
e-y = m_{14}^2 v_1 v_4 &,& f = m_{34}^2 v_3 v_4 \, .
\label{eq:corresp}
\ea
Using the tri-triangle we can eliminate
the cosines of the phases $\delta_i$
in Eqs. (\ref{eq:minvevs})
in terms of the sides:
\ba
v_1 \frac {\partial V} {\partial v_1} & = &
    m_1^2 v^2_1
    + ( -{{a b} \over c} + {{x e} \over d} - {{y z} \over f})
 +  v^2_1 g({\bf v}) = 0 \nonumber\\
v _2 \frac {\partial V} {\partial v_2} & = &
    m_2^2 v^2_2
    + ( -{{a c} \over b} + {{d x} \over e})
 -  v^2_2 g({\bf v}) = 0 \nonumber\\
v_3 \frac {\partial V} {\partial v_3} & = &
    m_3^2 v^2_3
    + ( -{{b c} \over a} - {{z f } \over y})
 +  v^2_3 g({\bf v}) = 0 \nonumber\\
v_4 \frac {\partial V} {\partial v_4} & = &
    m_4^2 v^2_4
    + ( {{e d} \over x} - {{y f} \over z})
 -  v^2_4 g({\bf v}) = 0 \, .
\label{eq:tritvevs}
\ea
To solve the four equations above and find the
VEVs $v_i$ we need first to express the combinations of sides
in (\ref{eq:tritvevs}) in terms of the masses and VEVs
in Eq. (\ref{eq:corresp}).
The VEVs would give us the sides of the tri-triangle
and from them we would construct
the tri-triangle and read the angles $\delta_i$.
This would complete
the search for the CP violating minimum of the
scalar potential.

As we will prove in the next section,
no such solution for the VEVs $v_i$ can be found
without fine tuning the mass
parameters. This is because supersymmetry
and the gauge symmetries dictate
a too restricted form for the scalar potential
(a simple D-term contribution and no trilinear terms).
In the rest of the
paper we show explicitly why this is so and point out minimal
modifications that would induce nontrivial phases.

\section{The minimum}

To solve (\ref{eq:tritvevs}) we need to express the combinations involving
the nine sides of the tri-triangle in terms of masses and VEVs ({\it i.e.},
in terms of the six distances in Eq. (\ref{eq:corresp})). This
requires solving a 4th order equation; we simplify this procedure
by making a redefinition of the Higgs fields.
We rewrite the scalar potential for the neutral fields
in matrix notation:
\ba
V & = &
\left (
\begin{array}{cc}
\phi_1^\dagger  &  \phi_3^\dagger   \\
\end{array}
\right )
\left (
\begin{array}{cc}
m_1^2 &  m_{13}^2 \\
m_{13}^2 &  m_{3}^2 \\
\end{array}
\right )
\left (
\begin{array}{c}
\phi_1 \\
\phi_3 \\
\end{array}
\right ) +
\left (
\begin{array}{cc}
\phi_2^\dagger  &  \phi_4^\dagger   \\
\end{array}
\right )
\left (
\begin{array}{cc}
m_2^2 &  m_{24}^2 \\
m_{24}^2 &  m_{4}^2 \\
\end{array}
\right )
\left (
\begin{array}{c}
\phi_2 \\
\phi_4 \\
\end{array}
\right ) \nonumber\\
 & + & [ \left (
\begin{array}{cc}
\phi_1 &  \phi_3 \\
\end{array}
\right )
\left (
\begin{array}{cc}
m_{12}^2 &  m_{14}^2 \\
m_{32}^2 &  m_{34}^2 \\
\end{array}
\right )
\left (
\begin{array}{c}
\phi_2 \\
\phi_4 \\
\end{array}
\right ) + h.c. ]\nonumber\\
 & + &
{1 \over 8} (g^2+g'^2) [
\left (
\begin{array}{cc}
\phi_1^\dagger  &  \phi_3^\dagger   \\
\end{array}
\right )
\left (
\begin{array}{c}
\phi_1 \\
\phi_3 \\
\end{array}
\right ) -
\left (
\begin{array}{cc}
\phi_2^\dagger  &  \phi_4^\dagger   \\
\end{array}
\right )
\left (
\begin{array}{c}
\phi_2 \\
\phi_4 \\
\end{array}
\right ) ]^2 \, .
\ea
The first two matrices above can be diagonalized
through two unitary transformations (in our case two
rotations, since the mass matrices are real and symmetric)
of the neutral scalar fields:
$\left (
\begin{array}{c}
\phi'_1 \\
\phi'_3 \\
\end{array}
\right ) =
{\bf U}_1
\left (
\begin{array}{c}
\phi_1 \\
\phi_3 \\
\end{array}
\right )$;
$\left (
\begin{array}{c}
\phi'_2 \\
\phi'_4 \\
\end{array}
\right ) =
{\bf U}_2
\left (
\begin{array}{c}
\phi_2 \\
\phi_4 \\
\end{array}
\right )
$.
Note that the quartic term in the potential will not change
its form and
can be obtained just by replacing unprimed by primed fields.
Then, without loss of generality we can go to a basis where
$m'^2_{13}=m'^2_{24}=0$ (from now on we drop the prime
to specify rotated quantities).
In Fig. 1 this corresponds to a tri-triangle where the sides
$1/d$, $1/e$, and $1/x$ disappear (that triangle
becomes infinite) and the quantities $b$ and $z$
become equal. Such an object depends on four independent
distances and contains three angles that vary (for different
choices of the distances) between 0 and $\pi$.
The minimum conditions can be
immediately read from Eq.(\ref{eq:corresp}) for this case:
\ba
a = m_{12}^2 v_1 v_2 &,& c = m_{32}^2 v_2 v_3 \, ,\nonumber\\
y = -m_{14}^2 v_1 v_4 &,& f = m_{34}^2 v_3 v_4 \, .
\label{eq:correspspec}
\ea
It is easy to see that if all the masses $m^2_{ij}$ in
Eq.~(\ref{eq:superpot})
are positive the absolute minimum will be real.
A necessary condition to have an absolute minimum
with complex phases is that
\be
m^2_{12}m^2_{14}m^2_{32}m^2_{34} < 0 \, .
\label{eq:product}
\ee
The signs of the masses $m^2_{ij}$ in Eq.~(\ref{eq:correspspec})
satisfy that constraint. As long as the choice of signs of
the mass terms satisfies (\ref{eq:product}),
the minimum can be obtained from
the above tri-triangle just by redefining the angles
$\delta_i$.

Now we try to solve the minimum conditions (\ref{eq:tritvevs})
which correspond to this particular tri-triangle.
The quantity $b$ is not an independent distance, the
solution can be determined in
terms of $a$, $c$, $f$, and $y$ (given in Eq.~(\ref{eq:correspspec}))
in Figure 1. We find
\be
{1\over b}  =  {1 \over {v_1 v_3}} h({\bf v}) \, ,
\ee
where
\be
h({\bf v}) = \sqrt{ m^2_{12} m^2_{34} - m^2_{14} m^2_{32}}
\sqrt{ \frac
{ {1 \over {m^2_{32} m^2_{34}}} v^2_1 - {1 \over {m^2_{12} m^2_{14}}} v^2_3 }
{ m^2_{12} m^2_{32} v^2_2 -  m^2_{14} m^2_{34} v^2_4 }  } \, .
\ee
Then the conditions (\ref{eq:tritvevs}) read
\ba
v_1 \frac {\partial V} {\partial v_1} & = & v^2_1
     \, [ \, m_1^2
 -{ { m^2_{12} m^2_{34} - m^2_{14} m^2_{32} } \over { m^2_{32} m^2_{34} } }
{1 \over h({\bf v})}  +  g({\bf v}) \, ] = 0 \nonumber\\
v _2 \frac {\partial V} {\partial v_2} & = &  v^2_2
    \, [ \, m_2^2
    - m^2_{12} m^2_{32} h({\bf v})
  - g({\bf v}) \, ] = 0 \nonumber\\
v_3 \frac {\partial V} {\partial v_3} & = &  v^2_3
    \, [ \, m_3^2
 +{ { m^2_{12} m^2_{34} - m^2_{14} m^2_{32} } \over { m^2_{12} m^2_{14} } }
{1 \over h({\bf v})}  + g({\bf v}) \, ] = 0 \nonumber\\
v_4 \frac {\partial V} {\partial v_4} & = &   v^2_4
    \, [ \, m_4^2
    + m^2_{14} m^2_{34} h({\bf v})
 - g({\bf v}) \, ] = 0 \, .
\label{eq:tritvevtwo}
\ea
Since these four equations depend only on two combinations
of VEVs, namely $g({\bf v})$
and $h({\bf v})$,
they are uncompatible (unless a fine tuned value of
the masses is chosen). We conclude that no
tri-triangle like solution exists for the tree level potential
in Eq.~(\ref{eq:vevpot}), and the three phases $\delta_i$ of the minima
are necessarily zero or $\pi$. This situation is somewhat similar
to the no-go theorem for simple singlet models \cite{roma86}.
In the tree level four Higgs model, SCPV
is not possible because of the specific form of the supersymmetric
potentials.

\section{Modifications of the model}

We find two types of modifications or additions that would change
this result, allowing for CP violating minima.
The first possibilty would require to add to the effective potential
new terms in order to avoid the cancellations producing that
only two combinations of the four VEVs appear in conditions
(\ref{eq:tritvevtwo}).
Since we already considered the most generic Lagrangian
consistent with (softly broken) SUSY and gauge invariance,
one is only left (of course, unless new fields are
introduced) with effective radiative corrections from
order one Yukawa couplings. In order to change at least two
of the four conditions
(\ref{eq:tritvevtwo})
 one needs two strong couplings. These
could be the two couplings of the top quark to $H_2$ and
$H_4$ or  a top coupling combined with a bottom coupling
to $H_1$ or $H_3$\footnote{
Note that since now the $Z$ mass has contributions
from four VEVs, it is possible to have order one bottom
Yukawa couplings even for order one
$\tan \beta \equiv \sqrt{v_2^2+v_4^2\over v_1^2+v_3^2}$.}.
Notice that in the second case the minimum conditions (\ref{eq:minphases})
will keep the same form as in the tree level case, and therefore the
tri-triangle solution works also here.
In order to suppress FCNC in these cases the simplest idea is to invoke some
additional symmetry that will suppress the Yukawa couplings
of the additional Higsses ($H_3$ and $H_4$).
Thus the second case, with
one top and one bottom Yukawa coupling large,
seems preferred.

A second possibility to obtain complex minima which does
not rely on radiative effects would imply a substantial
(but, in our opinion, well motivated) change on the definition
of the model. We have assumed that all couplings in the
potential are real and all phases are generated
spontaneously.
It seems plausible, however, to relax
this requirement and introduce {\it soft} CP violation
in the $\mu$ terms (see Eq.~(\ref{eq:pot})).
These terms could have their origin in some higher
scale (singlet VEVs) with no effects on
the rest of parameters.
It turns out that one can absorb three of the four $\mu$
phases by Higgs field redefinitions, resulting into a new
Lagrangian with only one complex phase $\delta_5$\footnote{This would
{\it break} one of the equations of (\ref{eq:minphases}) and destroy
the tri-triangle solution. However, since
the trivial CP conserving solution is also destroyed, the
{\it only} solution is automatically CP violating.}.
The origin
of CP violation would not be entirely spontaneous, but it would not
appear neither in an {\it uncontrolled} way. This scenario
seems more flexible for a treatment of FCNC in terms of an
additional symmetry\cite{rasi95,masi95} since no {\it a priori}
conditions on the sizes of the Yukawa couplings have been set.
It also avoids the typical domain wall problems of theories
with spontaneous breaking of discrete symmetries.

\section{Conclusions}

The models with four Higgs doublets are another
well motivated minimal extension of the MSSM.
We have studied the possibility of
SCPV in these models.
We found a simple geometrical
interpretation of the minimum equations that allowed us to understand
the conditions for SCPV.
Although no complex minima of the tree-level
potential are possible, we singled out two interesting possibilities
(radiative effects and explicit $\mu$ phases) that
modify the potential so that CP violating phases appear.
A more detailed analysis of these possibilities will be
discussed elsewhere \cite{masi95}.

\vspace{0.5in}

We thank R. Mohapatra for helpful suggestions and comments.
The work of M. M. has been partially supported by
a grant from the
Junta de Andaluc\'\i a (Spain).
The work of A. R. was supported by the NSF grant No. PHY9421385.
A. R. would like to thank G. Anderson, J. Kim, A. Melfo,
M. Primc and G. Senjanovi\'{c}
for discussions on solutions to the minimum equations.

\newpage
\textwidth 5.75in
\unitlength=1.00mm
\thicklines
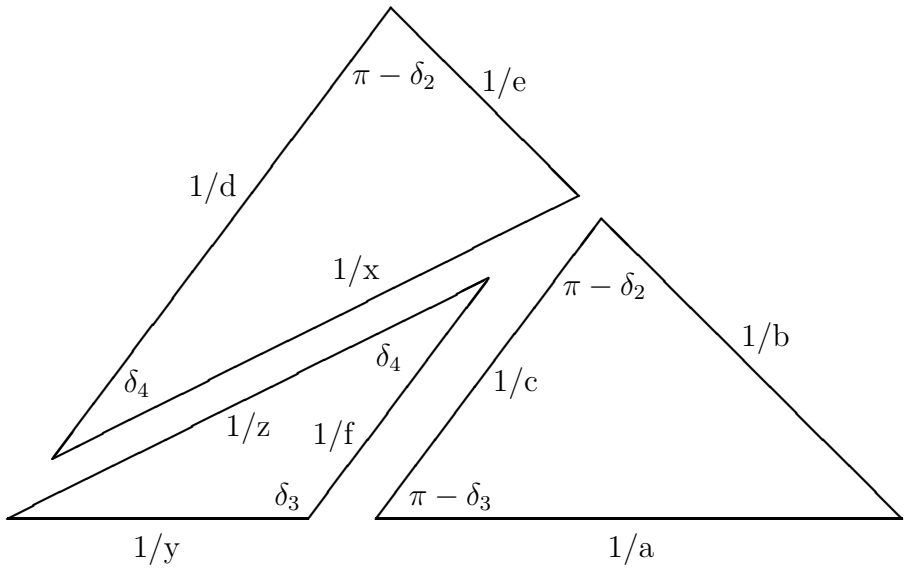
\begin{figure}
\begin{picture}(33.66,182.34)
\put(10.00,100.00){\line(1,0){40.00}}
\put(59.00,100.00){\line(1,0){70.00}}
\put(61.00,168.00){\line(-3,-4){45.00}}
\put(59.00,100.00){\line(3,4){30.00}}
\put(129.00,100.00){\line(-1,1){40.00}}
\put(61.00,168.00){\line(1,-1){25.00}}
\put(50.00,100.00){\line(3,4){24.00}}
\put(10.00,100.00){\line(2,1){64.00}}
\put(86.00,143.00){\line(-2,-1){70.00}}
\put(67.00,159.00){\makebox(0,0)[rc]{$\pi-\delta_2$}}
\put(95.00,131.00){\makebox(0,0)[rc]{$\pi-\delta_2$}}
\put(74.50,103.00){\makebox(0,0)[rc]{$\pi-\delta_3$}}
\put(49.00,103.00){\makebox(0,0)[rc]{$\delta_3$}}
\put(62.50,121.50){\makebox(0,0)[rc]{$\delta_4$}}
\put(29.00,118.00){\makebox(0,0)[rc]{$\delta_4$}}
\put(96.00,96.00){\makebox(0,0)[rc]{1/a}}
\put(114.00,124.00){\makebox(0,0)[rc]{1/b}}
\put(80.50,118.00){\makebox(0,0)[rc]{1/c}}
\put(40.50,144.00){\makebox(0,0)[rc]{1/d}}
\put(79.00,158.00){\makebox(0,0)[rc]{1/e}}
\put(56.00,111.00){\makebox(0,0)[rc]{1/f}}
\put(59.50,133.00){\makebox(0,0)[rc]{1/x}}
\put(33.00,96.00){\makebox(0,0)[rc]{1/y}}
\put(45.00,112.00){\makebox(0,0)[rc]{1/z}}
\end{picture}
\vspace{-8cm}
\caption{ The geometrical object which
represents the CP nontrivial solution of equations (9)
consists of three triangles. Each triangle contains two of the
three angles $\delta_2$, $\delta_3$ and $\delta_4$. The sides
of the triangles are denoted by 1/a, 1/b,
1/c, 1/d, 1/e, 1/f, 1/x, 1/y, 1/z.}
\end{figure}

\end{document}